\documentclass[twocolumn,superscriptaddress,prb,longbibliography,floatfix]{revtex4-2}
\usepackage{amsmath, amssymb}
\usepackage{graphicx}
\usepackage{times}
\usepackage{dcolumn}
\usepackage{multirow}
\usepackage{float}
\usepackage{graphicx}
\usepackage[colorlinks, linkcolor=blue, anchorcolor=blue, citecolor=blue]{hyperref}
\usepackage{verbatim}
\usepackage{autobreak}
\allowdisplaybreaks

\graphicspath{{figure/}{logo/}} 
\newcommand{\msr}{$\rm \mu$SR}

\newcommand{\Lamather}{La$_{2}$Cu$_{5}$As$_{3}$O$_{2}$}
\newcommand{\LaCuNiAsO}{La$_{2}($Cu$_{1-x}$Ni$_{x}$)$_{5}$As$_{3}$O$_{2}$}

\begin{document}

	\title{Superconducting Properties of \LaCuNiAsO: A \msr \ Study}
	\author{Qiong Wu}
	\author{Kaiwen Chen}
	\author{Zihao Zhu}
	\author{Cheng Tan}
	\author{Yanxing Yang}
	\author{Xin Li}
	\affiliation{State Key Laboratory of Surface Physics, Department of Physics, Fudan University, Shanghai 200438, People's Republic of China}
	\author{Toni Shiroka}
\affiliation{Laboratory for Muon-Spin Spectroscopy, Paul Scherrer Institute, CH-5232 Villigen PSI, Switzerland}
\affiliation{Laboratorium f\"ur Festk\"orperphysik, ETH Z\"urich, CH-8093 Z\"urich, Switzerland}
\author{Xu Chen}
 \author{Jiangang Guo}
	\author{Xiaolong Chen}
	\affiliation{Beijing National Laboratory for Condensed Matter Physics, Institute of Physics, Chinese Academy of Sciences, P. O. Box 603, Beijing 100190, China}
	\author{Lei Shu}
	\altaffiliation[Corresponding Author: ]{leishu@fudan.edu.cn}
	\affiliation{State Key Laboratory of Surface Physics, Department of Physics, Fudan University, Shanghai 200438, People's Republic of China}
	\affiliation{Collaborative Innovation Center of Advanced Microstructures, Nanjing 210093, People's Republic of China}
	\affiliation{Shanghai Research Center for Quantum Sciences, Shanghai 201315, People's Republic of China}
	\date{\today}
	
\begin{abstract}
We report the results of muon spin rotation and relaxation (\msr) measurements on the recently discovered layered Cu-based superconducting material \LaCuNiAsO \ ($x =$ 0.40, 0.45). Transverse-field \msr \ experiments on both samples show that the temperature dependence of superfluid density is best described by a two-band model. The absolute values of zero-temperature magnetic penetration depth $\lambda_{\rm ab}(0)$ were found to be 427(1.7) nm and 422(1.5) nm for $x =$ 0.40 and 0.45, respectively. Both compounds are located between the unconventional and the standard BCS superconductors in the Uemura plot. No evidence of time-reversal symmetry (TRS) breaking in the superconducting state is suggested by zero-field \msr \ measurements.

\end{abstract}
	
\maketitle
	
\section{Introduction}

The relation between magnetism and superconductivity is one of the most prominent issues in condensed matter physics~\cite{Luke1993,Kirtley2000,Kivelson2003,Balatsky2006,Onari2009,Hanaguri2010, Wu2011, Guguchia2014,Tarapada2020, Hayes2021}. Since the discovery of cuprate high transition temperature ($T_{\rm c}$) superconductors, considerable efforts have been made to investigate the role of in-plane impurities in them. It is now well established that, in copper oxide superconductors, nonmagnetic Zn ions suppress $T_c$ even more strongly than magnetic Ni ions~\cite{Nachumi1996,Smith2001, Yang2013, Guguchia2017}. Such behavior is in sharp contrast to that of conventional BCS superconductors, in which magnetic impurities can act as pairing-breaking agents, rapidly suppressing superconductivity~\cite{BTM1959,Anderson1959}. Another interesting behavior of unconventional superconducting systems such as the heavy-fermion, high $T_c$ cuprate, and iron-pnictide superconductors is the dome shape of the chemical doping dependence of $T_{\rm c}$~\cite{Li_2009,Stewart2011,Ideta2013,Dai2015}.

Recently, the first Cu-As superconductor was discovered in the layered La$_2$Cu$_5$As$_3$O$_2$ with $T_{\rm c} =0.63$~K~\cite{CHEN2019171}. When Cu$^{2+}$ is replaced by Ni$^{2+}$, also in \LaCuNiAsO, the $T_{\rm c}$ exhibits a dome-like structure. Remarkably, while the superconductivity in cuprate- and iron-based superconductors is completely suppressed when the substitution ratio of Cu or Fe exceeds 20\%~\cite{Wu1996, Itoh2002}, superconductivity in \LaCuNiAsO\ persists until the substitution ratio exceeds 60\%~\cite{CHEN2019171}. In this case, the robustness of superconductivity reveals the unexpected effect of impurities on inducing and enhancing superconductivity. Hence, \LaCuNiAsO \  provides a broader platform for studying the doping effect in the superconducting phase diagram.

Specific heat measurements have revealed that the optimally doped \LaCuNiAsO \ ($x =$ 0.40) sample shows a sharp superconducting transition at $T_{\rm c} =$ 2.05 K, with a dimensionless jump $C_{\rm e}/ \gamma_{\rm s} T_{\rm c} = 1.42$~\cite{CHEN2019171}, consistent with the BCS weak-coupling limit (1.43). In the superconducting state, the temperature dependence of the specific heat coefficient is described by a fully-gapped model $C_{\rm e} /T \propto e^{-\Delta/ k_{\rm B}T}$ after subtracting the upturn of $C_{\rm e}/T$ below $T < 0.5$~K, which is attributed to the Schottky effect. These results suggest that \LaCuNiAsO \ ($x =$ 0.40) is a conventional BCS superconductor with a fully developed energy gap. However, the fit yields $2\Delta / k_{\rm B} T_{\rm c}=2.58$~\cite{CHEN2019171}, much smaller than the BCS weak coupling limit. 

\msr \ experiments have been widely utilized to probe superconductivity in type-II superconductors at the microscopic level~\cite{AD2022}, and they are free from the influence of the Schottky effect. Transverse-field (TF) \msr \  measures the absolute value of the magnetic penetration depth $\lambda$, which is related to the density of superconducting carriers. The temperature dependence of $\lambda$ is sensitive to the lowest-lying superconducting excitations and provides information on the symmetry of superconducting pairing~\cite{Alain2010,Sonier2011,Zhang2016,Jian2018,Tan2018,Zhu_2022}. In addition, zero-field (ZF) \msr \ is a powerful method to detect small spontaneous internal magnetic fields due to the possible breaking of time-reversal symmetry (TRS) at the superconducting transition. These can be as small as 10~$\mu$T, corresponding to about $10^{-2}$ of Bohr magneton $\rm{ \mu_B}$~\cite{ Hayano1979, Amato1997, Aoki2003, Alain2010, Neha2019,AD2022}. 

Here, in order to study the doping effect on superconductivity, we perform \msr \ measurements on the polycrystalline samples of \LaCuNiAsO \  for $x = 0.40$ (optimal doping) and $x=0.45$ (overdoped). The temperature dependence of superfluid density determined from TF-\msr \ is best described by a two-band superconductivity model. $d$-wave superconductivity possibly exists in one of the bands, with the fraction of $d$-wave smaller in the overdoped sample than that in the optimally doped one. The superconducting energy gap is larger for optimal doping, suggesting that the coupling strength decreases with the increase of doping. 
Both compounds are located between the unconventional and the standard BCS superconductors in the Uemura plot. Meanwhile, no evidence of TRS breaking is suggested by ZF-\msr \ measurements.

\section{Experimental Details}

\label{Experiments}
	
Solid-state reactions were used to produce polycrystalline samples of \LaCuNiAsO \ ($x$ = 0.40, 0.45)  \cite{CHEN2019171}. The performed X-ray diffraction (XRD) studies and the density functional theory (DFT) calculation shows the band structures of \Lamather,  indicating that this newly synthesized sample is a layered superconducting compound, with the superconducting atomic layers consisting of the $[\rm{Cu_{5}As_{3}}]^{2-}$ cage-like structure \cite{CHEN2019171}.

\msr \ experiments were carried out using nearly 100\% spin-polarized positive muons ($\mu^+$) on the M15 beam line at TRIUMF, Vancouver, Canada for $x = 0.40$, and the DOLLY spectrometer of the S$\mu$S muon source at Paul Scherrer Institute (PSI), Switzerland for $x = 0.45$, respectively. The samples were mounted on a silver sample holder at TRIUMF and a copper sample holder at PSI, respectively. Only a very limited amount of muons stopped in the extremely thin copper sample holder.  In TF-\msr \ measurements, where the external field is applied perpendicular to the initial muon spin polarization, muons are implanted one at a time into a sample which is cooled (from above $T_c$) in an external magnetic field. Muon spins precesses around the local field at the implantation site, and the functional form of the muon spin polarization depends on the field distribution of the vortex state, including the magnetic penetration depth, the vortex core radius, and the structure of the flux-line lattice. For ZF-\msr \ measurements, the ambient magnetic field was actively compensated to better than 1 $\mu$T. \msr \ data were analyzed using the \textsc{musrfit} software package \cite{SUTER201269}.

\section{RESULTS}
	
\subsection{Transverse-field \msr \ experiments}
	
\begin{figure}[!ht]
	\centering
	\includegraphics[clip=,width=8.6cm]{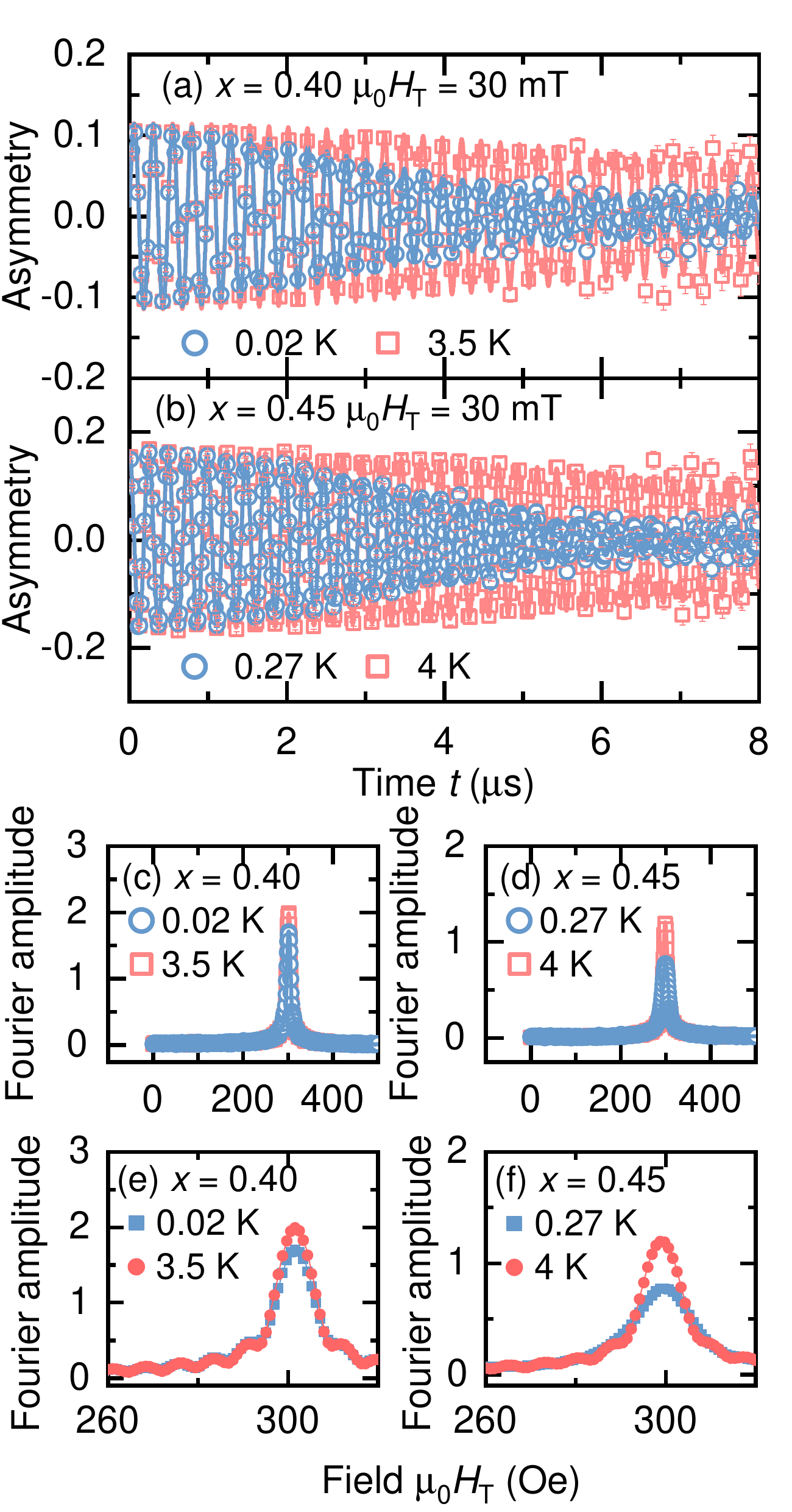}
	\caption{(a)-(b) Representative TF-\msr \ asymmetry spectra $A(t)$ for \LaCuNiAsO \ ($x=$ 0.40 (panel (a)) and 0.45 (panel (b))) in the normal (red squares) and superconducting (blue circles) states with an external magnetic field of $\rm{\mu_0}H =$ 30 mT after subtracting the background signal. Solid curves: fits to the data using Eq.~(\ref{TF Equation}). (c)-(f) Fourier transformation of the total TF-\msr \ time spectra. (e)-(f) show the details of the FFT spectra.
}
	\label{FIG1}
\end{figure}

The \msr \ asymmetry spectrum usually consists of a signal from muons that stop in the sample and a slowly relaxing background signal from muons that miss the sample for example, stop in the sample holder. Fig.~\ref{FIG1} (a) (b) show the typical TF-\msr \ muon spin precession signals at an applied field of $\mu_{\rm{0}}H =$ 30~mT in the normal (red squares) and superconducting states (blue circles) for \LaCuNiAsO \ ($x =$ 0.40 and 0.45) after subtracting the background signal. The superconducting volume fractions are estimated to be 60$\%$ and 70$\%$ from the TF \msr \ asymmetry at long times for x = 0.40 and 0.45, respectively.
Fig.~\ref{FIG1} (c) (d) show the Fourier transformations (FFT) of the total TF-\msr \ asymmetry. It can be inferred from the figure that vortex states are constructed in both samples at low temperatures \cite{Karl2019}.
Fig.~\ref{FIG1} (e) (f) show the details of the FFT spectra. As shown in the figure, the magnetic fields in the superconducting state and the normal state are relatively close to each other. This is common in anisotropic powder superconducting samples \cite{Weber1993,Maisuradze_2009}.

The TF-\msr \ time spectra after subtracting the background signal can be well fitted by the function
\begin{equation}
	\label{TF Equation}
	A(t)=A_0\exp(-\frac{1}{2}{\sigma_{\rm TF}^2t}^2)\cos(\gamma_{\rm \mu} B_{\rm s}t + \varphi_{s}), 
\end{equation}
where A$_0$ is the initial asymmetry of the muon spin in the sample.
The Gaussian relaxation rate $\sigma_{\rm TF}$ due to the nuclear dipolar fields in the normal state is enhanced in the superconducting state due to the field broadening generated by the emergence of the flux-line lattice (FLL). $\gamma_{\rm \mu} = 8.516$ $\times$ 10$^8$~s$^{-1}$T$^{-1}$ is the gyromagnetic ratio of the muon, and $B_{\rm s}$ is the magnetic field at muon stopping sites. 

\begin{figure}[ht]
	\centering
	\includegraphics[clip=,width=8.6cm]{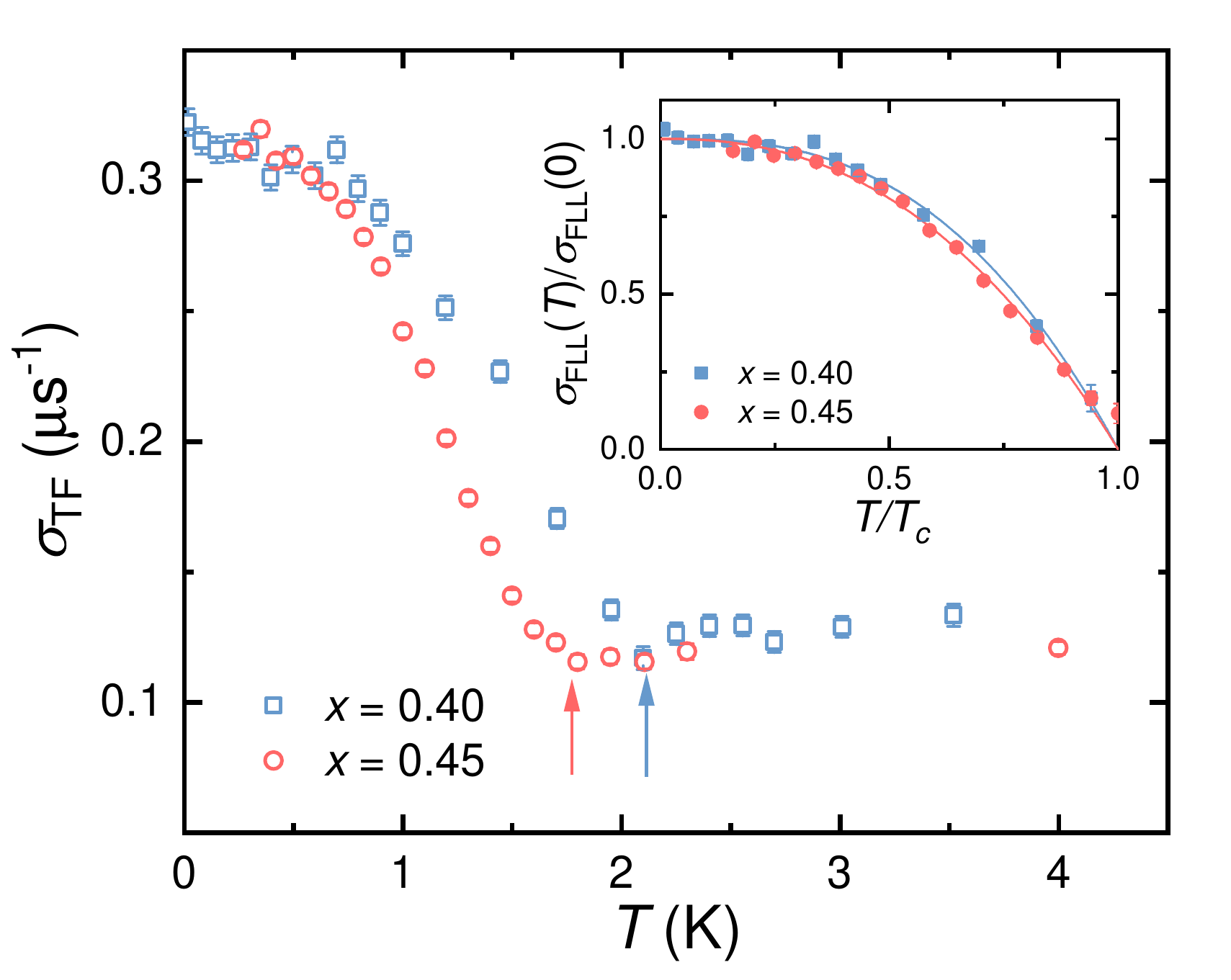}
	\caption{Temperature dependencies of TF-\msr \ Gaussian relaxation rate $\sigma_{\rm TF}$ in \LaCuNiAsO \ ($x = 0.40$ and 0.45) for $\rm{\mu_{0}}$$H = 30$ mT. The arrows indicate $T_{\rm c}$ from resistivity measurements \cite{CHEN2019171}. Inset: the normalized superfluid density $\sigma_{\rm FLL}(T)/\sigma_{\rm FLL}(0)$ from Eq.~(\ref{sigmaTFsc}) vs. the reduced temperature $T/T_{\rm c}$. Solid curves: phenomenological two-fluid model fits (see text).}
	\label{sigmaTF}
\end{figure}

Figure~\ref{sigmaTF} shows the temperature dependence of $\sigma_{\rm TF}$ obtained from the fits using Eq.~(\ref{TF Equation}) for \LaCuNiAsO \ ($x =$ 0.40 and 0.45). There are noticeable upturns in $\sigma_{\rm TF}$ that develop below $T_c=2.2$~K and 1.8 K, respectively.
The internal field distribution in the vortex state is the convolution of the FLL field distribution and the nuclear dipolar field distribution of the host material:
\begin{equation}
	\label{sigmaTFsc}
	\sigma_{\rm TF}= \begin{cases}
		\sqrt{\sigma_{\rm FLL}^2+\sigma_{\rm dip}^2}  &  (T \leqslant T_{c})\\
		\sigma_{\rm dip} &  (T > T_{c})
	\end{cases}.
\end{equation} 

Above $T_{\rm c}$, $\sigma_{\rm TF}=\sigma_{\rm dip}$ generated from the nuclear dipolar fields is roughly independent of temperature. Therefore, it was fixed to the average value (0.1179(11) $\rm{\mu  s^{-1}}$ for $x = 0.45$  and 0.127(2) $\rm{\mu  s^{-1}}$ for $x = 0.40$). The value of $\sigma_{\rm dip}$ is smaller for larger $x$, consistent with the fact that the nuclear magnetic moment of nickel is smaller than that of copper \cite{Ashcroft1976}.
		
In the inset of Fig.~\ref{sigmaTF}, the normalized FLL relaxation rate $\sigma_{\rm FLL}(T)/\sigma_{\rm FLL}(0)$ is plotted vs. the reduced temperature $T/T_{\rm c}$ for $x =$ 0.40 and 0.45. The data can be fitted with the phenomenological two-fluid model~\cite{Hillier1997, Ding2019}
\begin{equation}
		\label{eq:powerlaw}
		\sigma_{\rm FLL}(T)=\sigma_{\rm FLL}(0)[1-(T/T_{\rm c})^N] \quad (T \leqslant T_{\rm c}).
\end{equation}
The fitting parameters are listed in Table~\ref{tab:powerlaw}. 
There are several common predictions for the value of $N$. For traditional BCS superconductors, $N \sim 4$. However, both values of $N$ for $x =$ 0.40 and 0.45 are much lower than 4. The dirty-limit $d$-wave model predicts a value of $N=2$ \cite{Luetkens2008, Hirschfeld1994}.
However, in the \LaCuNiAsO \ system, the upper critical field $H_{\rm c2}(0)$ is estimated to be 3 T \cite{CHEN2019171}, giving an estimated value of the Ginzburg-Landau coherence length $\xi(0)=[\Phi_{0}/2\pi H_{c2}(0)]^{1/2}=10.47$ nm, where $\Phi_0 =2.07\times{10}^{-3}$ Tm$^2$ represents the magnetic flux quantum. 
Given the metallic behavior, the mean free path can be estimated by $l_e=\hbar k_{\rm F}/\rho_0 ne^2$ \cite{Wang2022}, with residual resistivity $\rho_0=$0.34~$\rm{m\Omega cm}$ \cite{CHEN2019171}, yielding $l_e$ = 15 nm for $x=0.40$.
Thus, \LaCuNiAsO \ may be a relatively clean superconductor with $\xi/l_e \approx 1$.
The predicted clean-limit $d$-wave model $N$ value is $N=1$ \cite{Coleman2015}.
As a result, our system may include both $s$ and $d$ waves.

\begin{table}[H]
	\caption{\label{tab:powerlaw} Parameters from fits to TF-\msr \ data using the phenomenological two-fluid model for \LaCuNiAsO.}
	\begin{ruledtabular}
		\begin{tabular}{ccccc}
			Parameters  & $x = 0.40$ & $x = 0.45$\\
			\hline
			$\sigma_{\rm FLL}(0)~(\rm{\mu s^{-1}})$  & $0.287(2)$ & $0.300(4)$\\
			$T_{c}~(\rm{K})$    & $2.08(4)$ & $1.70(2)$ \\
			$N$    &$2.72(17)$ & $2.38(13)$ \\
			$\lambda_{\rm eff}(0) ~ (\rm nm)$ & $559.4(1.9)$ & $547(3)$\\
			$\lambda_{\rm ab}(0) ~ (\rm nm)$  & $427.02(1.49)$ & $418(3)$\\
			$\sigma_{\rm dip}~(\rm{\mu s^{-1}})$  & $0.127(2)$ & $0.1179(11)$\\
			Adj. R$^2$     & 0.98443 & 0.99172
		\end{tabular}
	\end{ruledtabular}
\end{table}

For powder superconductor samples with $0.13/\kappa^2 \ll H / H_{\rm c2} \ll 1$ \ ($\kappa=\lambda/\xi$ is the Ginzburg-Laudau parameter, $H$ is the applied field), the Gaussian depolarization rate $\sigma_{\rm FLL}$ is directly related to the magnetic penetration depth $\lambda_{\rm eff}$ by~\cite{Brandt1988,Brandt2003}
\begin{equation}
	\label{eq:lambda}	
	\begin{aligned}
		\frac{\sigma_{\rm FLL}}{\gamma_\mu}= \frac{0.172(1-b)[1+1.21(1-\sqrt{b})^{3}]\Phi_0}{2\pi\lambda_{\rm eff}^{2}},
	\end{aligned}
\end{equation}
where $b$ is the reduced applied field $b=H / H_{\rm c2} = 0.01\ll 1$.  Therefore, the absolute values of effective magnetic penetration depth $\lambda_{\rm eff}(0)$ can be obtained and listed in Table~\ref{tab:powerlaw}. In addition, for layered superconductors, the in-plane magnetic penetration depth $\lambda_{\rm ab}(0)$ is also estimated by the relation $\lambda_{\rm eff}(0) = 3^{1/4} \lambda_{ab}(0)$~\cite{Barford1988,Fesenko1991,Zhang2016}.

\begin{figure}[!ht]
	\centering
	\includegraphics[clip=,width=8.6cm]{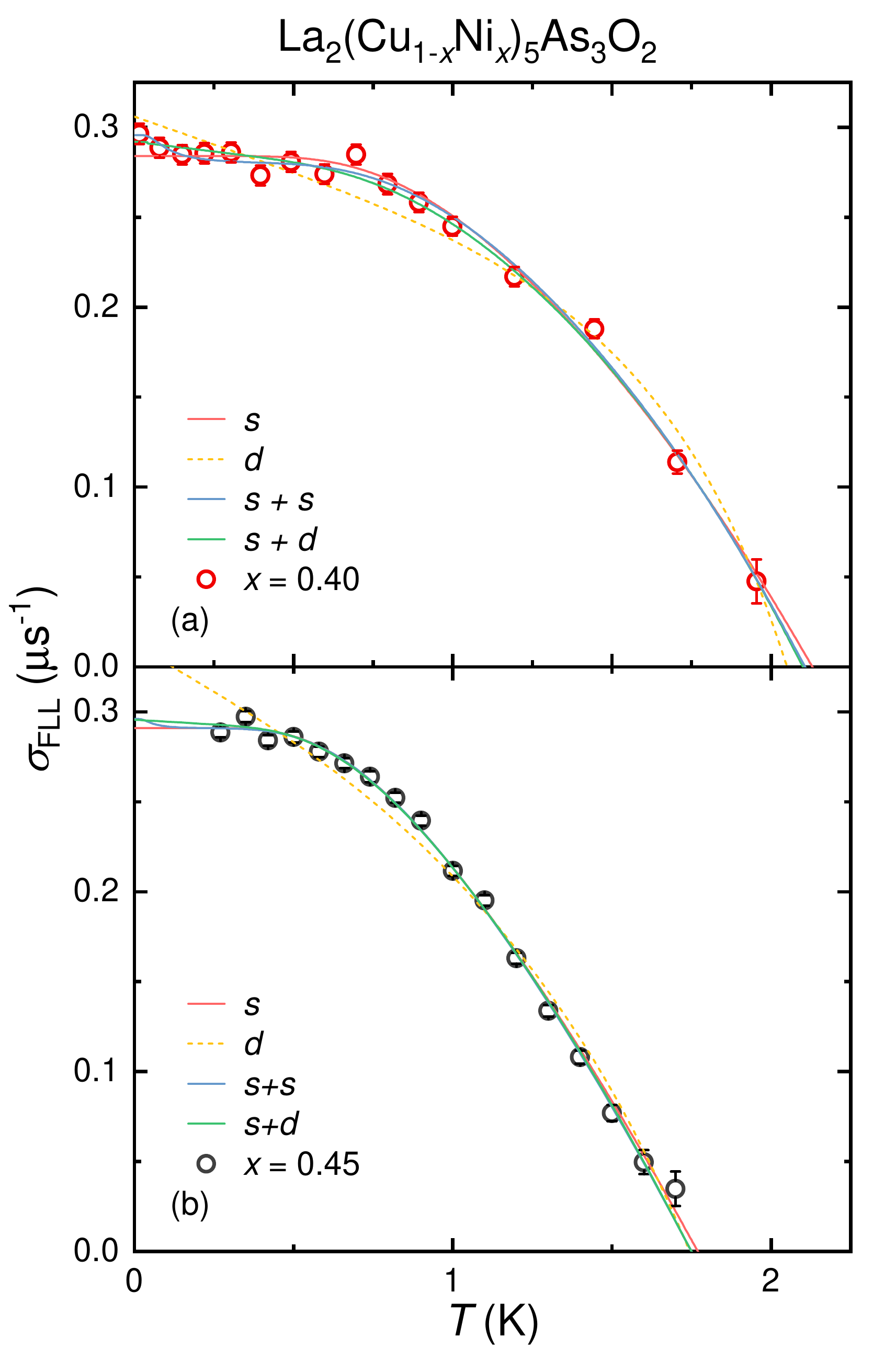}
	\caption{The FLL relaxation rate $\sigma_{\rm FLL}(T)$ from Eq.~(\ref{sigmaTFsc}) vs. temperature for \LaCuNiAsO \ , $x = 0.40$ (a) and 0.45 (b). Curves correspond to the $s$-wave, $d$-wave, $s+s$-wave, and $s+d$-wave models.}
	\label{FIG3}
\end{figure}

The effective magnetic penetration depth $\lambda_{\rm eff}$ for isotropic superconductors is related to the density $n_{\rm s}$ of superconducting carriers and $m^*$ by the general London equation~\cite{Hillier1997}
\begin{equation}
	\label{eq:London formula}
	\lambda_{\rm{eff}}^2=\frac{m^*}{\mu_0 e^2 n_{\rm s}},
\end{equation}
where $\mu_0$ is the magnetic constant, $m^*$ is the effective electron mass, and $e$ is the elementary charge. 
Therefore, the FLL relaxation rate $\sigma_{\rm FLL}$ is directly related to the superfluid density through $ \sigma_{\rm FLL} \propto \lambda_{\rm{eff}}^{-2} \propto n_{\rm s}$.
Then we can use microscopic models to investigate the gap symmetry of \LaCuNiAsO \ in more detail, as shown in Fig.~\ref{FIG3}.

\begin{table*}[h!t]
	\caption{\label{tab:temp} Parameters from the fits to the single-gap and two-gap models from TF-\msr \ data. The weighting factor of phenomenological two-gap $\alpha$ model $f_{\Delta_1}$, zero-temperature superconducting gap $\Delta(0)$, zero-temperature magnetic penetration depth $\lambda_{ab}(0)$, and reduced $\rm \chi_{red}^2$ from fits of Eq. (\ref{eq:gap}) and Eq. (\ref{eq:two-gap}). }
	\begin{ruledtabular}
		\begin{tabular}{cccccccccc}
			Ni doping & Model  & $f_{\Delta_1}$  & 2$\Delta_1/k_B T_c$  & 2$\Delta_2/k_B T_c$  & $\lambda_{ab}(0)$ (nm) & $\rm \chi^2_{red}$ \\
			\hline
			&	$s$    &    1       & 3.69(17)         &              & 427.4(1.8)       & 1.74    \\
			&	$d$      &   1      & 6.68(37)         &                & 411(3)       & 4.46    \\
			0.40 &	$s+s$   & 0.95(3)     & 3.86(17)        & 0.20(13)     & 419(5)       & 1.31    \\
			&	$s+d$      & 0.70(12)   & 4.03(21)       & 5.62(7)       & 422(3)       & 1.50    \\	    
			\hline
			&	$s$    & 1           & 3.34(7)        &          & 421.5(1.7)      & 2.19    \\
			&	$d$    & 1          & 5.09(26)         &           & 392(5)       & 12.9    \\
			0.45 &	$s+s$  & 0.988(154)       & 3.42(2)   & 0.262(2)   & 420(31)   & 2.06  \\
			& $s+d$  & 0.976(114)       & 3.43(3)       & 4.15(3)     & 422(5)      & 2.09   
		\end{tabular}
	\end{ruledtabular}
\end{table*}

The single-gap model is defined within the local London approximation \cite{2010FeSe}:
\begin{equation}
	\label{eq:gap}
	\frac{\sigma_{\rm FLL}(T)}{\sigma_{\rm FLL}(0)}=1+\frac{1}{\pi} \int_{0}^{2 \pi} \int_{\Delta(T, \varphi)}^{\infty}\frac{\partial f(E)}{\partial E} \frac{E\ \mathrm{d} E\ \rm{d} \varphi}{\sqrt{E^{2}-\Delta^{2}(T, \varphi)}},
\end{equation}
\begin{equation}
	\label{eq:Delta}
	\Delta(T, \varphi)=\Delta_{0} \delta\left(T / T_{\rm c}\right) g(\varphi),
\end{equation}
\begin{equation}
	\label{eq:gapT}
	\delta\left(T / T_{\rm c}\right)=\tanh \left\{1.82\left[1.018\left(T_{\rm c}/T-1\right)\right]^{0.51}\right\},
\end{equation}
where $\sigma_{\rm FLL}(0)$ is the FLL relaxation rate at zero temperature, $f(E)$ is the Fermi-Dirac distribution function, $\varphi$ is the angle along the Fermi surface, and $\Delta_{0}$ is the maximum superconducting gap value at $T = 0$. $g\left(\varphi\right)$ in Eq.~(\ref{eq:Delta}) describes the angular dependence of the superconducting gap. Here $g\left(\varphi\right)=1$ and $|\cos(2\varphi)|$ refers to the $s$-wave and $d$-wave model, respectively. 

The fitting parameters are listed in Table~\ref{tab:temp}. It is clear from Fig.~\ref{FIG3} that the $d$-wave model (the dashed yellow lines) does not fit the data, also evidenced by the largest $\chi^2_{\rm red}$ in Table~\ref{tab:temp}. 
The solid red curves in Fig.~\ref{FIG3} representing the $s$-wave model seem to fit the data well giving the reduced chi-square $\rm \chi^2_{red}$ 1.74 and 2.19 for $x$ = 0.40 and 0.45, respectively. Thus our results preliminarily suggest the $s$-wave pairing for both overdoped and underdoped samples for \LaCuNiAsO \ ($x$ = 0.45, 0.40). $\Delta(0)$ is larger for the optimal doping sample, suggesting the coupling strength decreases with the increase of doping concentration.

However, we notice that the superfluid density of $x =$ 0.40 has a minor upturn at low temperatures. This may be due to nodal or multiband superconductivity. As a result, in addition to single-gap functions, we also employ the phenomenological two-gap $\alpha$ model with a weighting factor $f_{\Delta_1}$ ~\cite{Padamsee1973, CARRINGTON2003, Khasanov2007},
\begin{equation}
	\label{eq:two-gap}
	\frac{\sigma_{\rm FLL}(T)}{\sigma_{\rm FLL}(0)}=f_{\Delta_1}	\frac{\sigma_{\rm FLL,\Delta_1}(T)}{\sigma_{\rm FLL,\Delta_1}(0)}+(1-f_{\Delta_1})	\frac{\sigma_{\rm FLL,\Delta_2}(T)}{\sigma_{\rm FLL,\Delta_2}(0)},
\end{equation}
where ${\sigma_{\rm FLL,\Delta_i}^{-2}(T)}/{\sigma_{\rm FLL,\Delta_i}^{-2}(0)} ~(i=1,~2)$ is the superfluid density contribution of one of the gaps.

While the smallest $\chi^2_{\rm red}$ in both compounds suggests that the $s+s$ model may be the best model, the resulting values of 2$\Delta_2/k_B T_c$ is unreasonably small. The $s+d$ model is also better than the $s$-wave model according to the $\chi^2_{\rm red}$ values. 
Moreover, the gap-to-$T_c$ ratios 2$\Delta_{1,2}/k_B T_c$ determined from the $s+d$ model are also close to the BCS theoretical predictions (2$\Delta_{s}/k_B T_c=$ 3.43(3) for $s$-wave, and 2$\Delta_{d}/k_B T_c=$ 4.15(3) for $d$-wave, respectively) for $x$ = 0.45. Both $s+s$ and $s+d$ models can well deal with the upwarping phenomenon of superfluid density at low temperatures in $x$ = 0.40. However, we also notice that the relaxation rate of $x$ = 0.45 does not show obvious upwarping at the current lowest temperature of 0.27 K. And also the obtained $f_{\Delta_d}$ values for both compounds are extremely small. More studies are needed to determine whether $d$-wave superconductivity exists in one of the bands.

\subsection{Uemura Plot}
	
\begin{figure}[!ht]
	\centering
	\includegraphics[clip=,width=8.6cm]{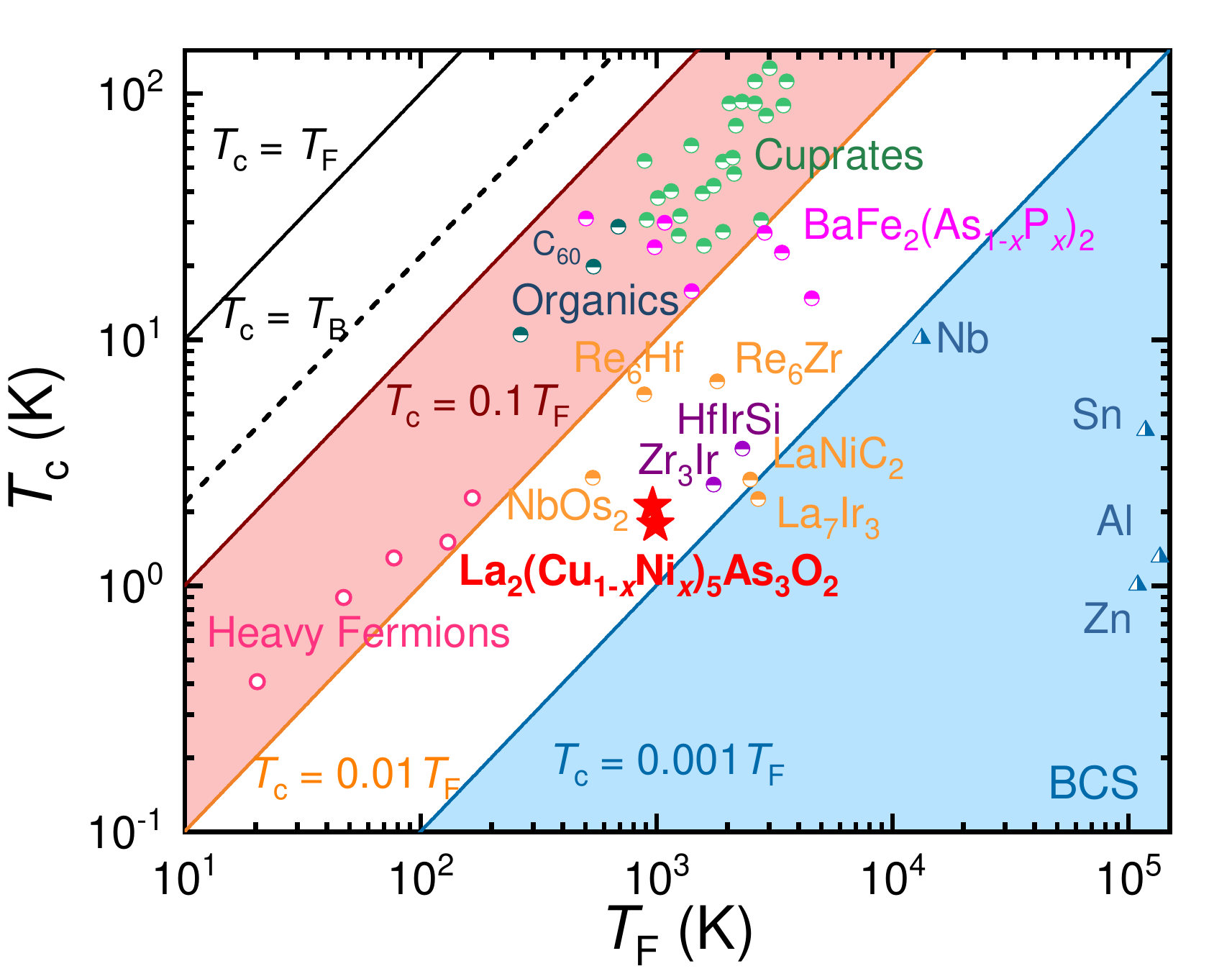}
	\caption{The Uemura classification scheme \cite{Uemura1989,Uemura2004,Barker2018}. The superconducting transition temperature $T_{\rm c}$ versus the effective Fermi temperature, $T_{\rm F}$, where the position of \LaCuNiAsO \ ($x =$ 0.45, 0.40) is shown by the red stars. The unconventional superconductors fall within a band indicated by the red and orange lines. The BCS superconductors are in the lower right region with blue marks.}
	\label{Uemurafig}
\end{figure}

An Uemura plot \cite{Uemura1989,Uemura2004,Yuji2021} is shown in Fig.~\ref{Uemurafig}, including elemental superconductors (such as Nb, Al, Sn, and Zn), cuprates, alkali-doped C$_{60}$ (K$_3$C$_{60}$ and Rb$_3$C$_{60}$), heavy-fermion superconductors, and \LaCuNiAsO. In this classification, unconventional superconductors usually lie in the orange area for $1/100 \leqslant T_{\rm c}/T_{\rm F} \leqslant 1/10$, conventional BCS superconductors fall in the blue region for $T_{\rm c}/T_{\rm F} \leqslant 1/1000$, where $T_{\rm F}$ is the Fermi temperature~\cite{Hillier1997,Uemura2004}.
	
For the quasi-2D systems, $T_{\rm F}$ can be estimated by the following relation~\cite{Hillier1997,Benfatto2001},
	\begin{equation}
		\label{eq:uemura plot}
		k_\mathrm{B}T_\mathrm{F}=\frac{\pi\hbar^2n_{\rm 2D}^s}{m^\ast},
	\end{equation}
where $n_{\rm{2D}}^s$ is the two dimensional superfluid density within the superconducting planes derived from the in-plane superfluid density via $n_{\rm{2D}}^s = n_{\rm{ab}}^sd$, where $d$ represents the interplanar distance. 
According to the general London equation Eq.~(\ref{eq:London formula}), we can get the in-plane superfluid density through $n_{\rm{ab}}^s = \frac{m^*}{\mu_0 e^2 \lambda_{\rm{ab}}^2}$.

Correspondingly, 
\begin{equation}
		\label{eq:Fermi temperaure}
		T_{\rm F}=\frac{\pi \hbar^2\lambda_{\rm{ab}}^{-2}d}{k_{\rm B}\mu_0 e^2},
\end{equation}
therefore, $T_{\rm F}$ of  \LaCuNiAsO \ ($x =$ 0.45 and 0.40) are estimated, and the results are listed in Table~\ref{Uemura}. As shown in Fig.~\ref{Uemurafig}, \LaCuNiAsO \ ($x =$ 0.40 and 0.45) fall in the crossover between the unconventional superconducting area and the conventional BCS region. The fact that $T_{\rm c}/T_{\rm F}$ is larger in $x =$ 0.40 than 0.45 may suggest a stronger pairing interaction in $x=0.40$. 

\begin{table}[H]
		\caption{\label{Uemura} Uemura plot parameters for \LaCuNiAsO }
		\begin{ruledtabular}
			\begin{tabular}{ccccccc}
				$x$ & $\lambda_{\rm{ab}}(0)$ (nm)   & $d$ ($\rm{\AA}$)  & $T_{\rm F}(\rm{K})$    & $T_{\rm c}(\rm{K})$  & $T_{\rm c}/T_{\rm F}$       \\
				\hline
				0.40               & 427 & 22.44 & 964 & 2.11  & $\sim$1/457 \\
				0.45              & 422 & 22.43 & 990 & 1.77  & $\sim$1/560
			\end{tabular}
		\end{ruledtabular}
\end{table}
	
\subsection{ZF-\msr}
	
\begin{figure}[!ht]
	\centering
	\includegraphics[clip=,width=8.6cm]{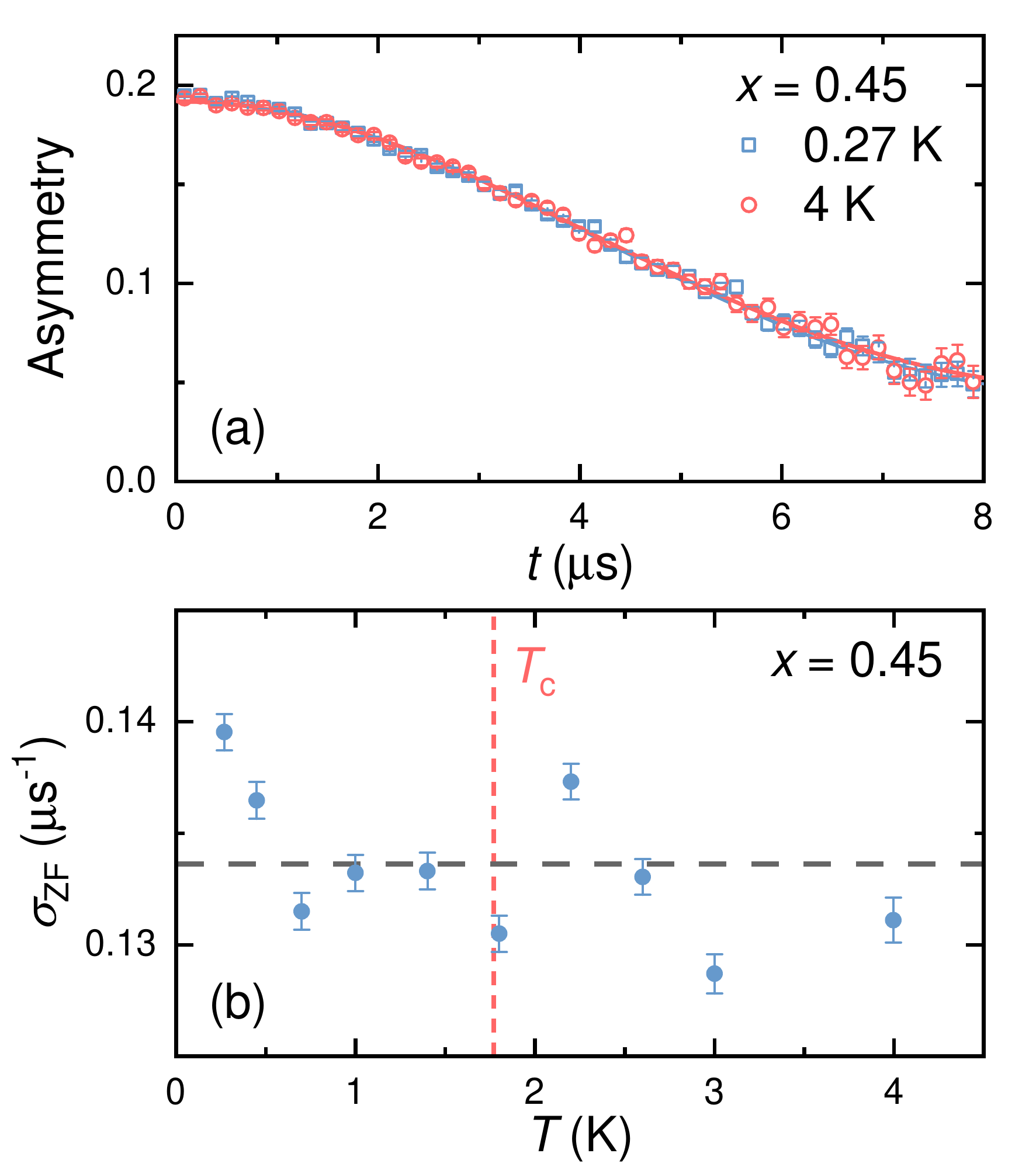}
	\caption{(a) ZF-\msr \ time spectra at representative temperatures for polycrystalline samples of \LaCuNiAsO \ ($x$ = 0.45). Curves: Fits to the data by Eq.~(\ref{eq:ZF function}). The background signal from muons that stop in the copper sample holder has not been subtracted. (b) Temperature dependence of the Gaussian KT relaxation rate $\sigma_{\rm ZF}$ for \LaCuNiAsO \ ($x$ = 0.45). 
	The red dash line represents $T_c$ determined from the transport measurements. 
	The gray dashed line shows an average of $\sigma_{\rm ZF}$ = 0.1336 $\mu$s$^{-1}$.}
	\label{fig:Asy}
\end{figure}

To further investigate the superconductivity in \LaCuNiAsO, ZF-\msr \ experiments were performed. Fig.~\ref{fig:Asy} (a) shows the time evolution of the decay positron count asymmetry, which is proportional to the muon spin polarization, at temperatures above and below $T_c$ in $x =$ 0.45. There is no noticeable difference between the superconducting and the normal state, suggesting the absence of a spontaneous magnetic field below $T_c$. Therefore, TRS is conserved below the superconducting transition. Over the entire temperature range, the ZF-\msr \ spectra can be well described by the following function,
\begin{equation}
		\label{eq:ZF function}
		\frac{A\left(t\right)}{A(0)}=(1-f_{\rm bg}) G_{\rm KT}(t)+ f_{\rm bg}G_{\rm bg}(t),
\end{equation}
where the two relaxing terms are the signals of the sample and the background. $A(0)$ is the initial total asymmetry at time $t =$ 0, and $f_{\rm bg}$ is the proportion of the background signal, which has the same value as the one in the TF-\msr \ experiments.  $G_{\rm KT}(t)$ is the static Gaussian Kubo-Toyabe (KT) function~\cite{Hayano1979}
\begin{equation}
		\label{eq:KT}
		G_{\rm KT}(\sigma_{\rm ZF},t)=\frac{1}{3}+\frac{2}{3}\left(1-\sigma^2_{\rm ZF} t^2\right)\exp\left(-\frac{1}{2}\sigma^2_{\rm ZF} t^2\right),
\end{equation}
where $\sigma_{\rm ZF}$ is the Gaussian KT relaxation rate which corresponds to the relaxation due to static, randomly oriented local fields associated with the nuclear moments at the muon site. Fig.~\ref{fig:Asy} (b) shows that there is no significant change in relaxation rate $\sigma_{\rm ZF}$ down to base temperature $T=0.27$~K. The average value of $\sigma_{\rm ZF}$ is 0.1336 $\mu s^{-1}$, consistent with the typical nuclear dipolar moments of Ni and Cu \cite{LukeGM1991,LaNiC2-2009,RPSingh2020}.

\section{DISCUSSION}

Low-temperature investigations are crucial for determining the superconducting pairing mechanism. Our TF-\msr \ measurements were performed down to 20 mK, and the temperature dependence of superfluid density suggests that the two-band model best fits the data for both measured samples, one of which is dominated by the $s$ wave and the other in such a small proportion that we cannot confirm whether it is the $s$ wave or the $d$ wave under the current data conditions.
The two-band superconductivity scenario is also supported by the density functional theory (DFT) calculations showing that two bands crossing the Fermi energy contribute to the Fermi surface in \Lamather~\cite{CHEN2019171}. 
Furthermore, the temperature dependence of the upper critical field $H_{c2}(T)$ of La$_{2}( $Cu$_{1-x}$Ni$_{x}$)$_{5}$As$_{3}$O$_{2}$ exhibits a upward curvature~\cite{CHEN2019171}, significantly different from the Werthamer-Helfand-Hohenberg relation~\cite{WHH1966}, which may also proposes multi-band superconductivity~\cite{Hunte2008}.

BCS superconductivity and Bose-Einstein condensation (BEC) are two asymptotic limits of a fermionic superfluid. Systems with a small $T_{\rm c}/T_{\rm F}$ \textless \ 0.001 are usually considered to be BCS-like, while large $T_{\rm c}/T_{\rm F}$ values are expected only in the BEC-like picture and are considered to be a hallmark feature of unconventional superconductivity~\cite{Uemura1989,Uemura1991,Hillier1997}. As shown in Fig. \ref{Uemurafig}, \LaCuNiAsO \ falls in the crossover area between BCS and exotic superconductors regions, like many other superconductors such as TRS breaking superconductors La$_7$Ir$_3$ and LaNiC$_2$~\cite{LaNiC2-2009,La7Ir3-2015,Barker2018}, 
and multi-band iron-based superconductors BaFe$_2$(As$_{1-x}$P$_x$)$_2$~\cite{BaFe2As2-2012}.
In the Uemura plot classification scheme, these superconductors may not be traditional electron-phonon coupled BCS superconductors but more like exotic unconventional superconductors. 

More experimental evidence is required to investigate whether $d$-wave superconductivity exists in one of the bands. Table~\ref{tab:temp} shows that the proportion of $d$-wave decreases as Ni concentration increases if it does exist. Such a behavior is rare but was observed in layered cuprates La$_{2-x}$Ce$_x$CuO$_{4-y}$ and Pr$_{2-x}$Ce$_x$CuO$_{4-y}$, where a transition from $d$- to $s$-wave pairing occurs near the optimal doping~\cite{Skinta2002}.
Also in LaFeAs$_{1-x}$P$_x$O~\cite{Shi2018}, the superconducting
order parameter evolves from nodal to nodeless as the doping concentration exceeds 50\%.
Moreover, a crossover from a nodal to nodeless superconducting energy gap was also suggested in skutterudite PrPt$_4$Ge$_{12}$ through Ce substitution~ \cite{Mai2009,Huang2014}, although the possibility of a transition from multi-band to single-band superconductivity can not be excluded. In addition, a molecular pairing scenario~\cite{Coleman2015} was proposed to explain the transition from nodal to nodeless superconductivity in Yb-substituted CeCoIn$_5$~\cite{CeCoIn5_TDO}. The Yb doping increases the chemical potential and drives a Lifshitz transition of the nodal Fermi surface, forming a fully gapped molecular superfluid of composite pairs. For \LaCuNiAsO, the Fermi pocket around the $\Gamma$ point is relatively small according to the DFT calculations~\cite{CHEN2019171}, and detailed electronic structure study is required to investigate whether the molecular pairing scenario can be applied.

\section{Conclusions}
	
In summary, we performed ZF and TF -\msr \ measurements on the recently discovered layered superconductor \LaCuNiAsO \ ($x =$ 0.40, 0.45). The preservation of TRS is suggested by the ZF-\msr \  measurements. In combination with the $H_{c2}(T)$ data and DFT calculations \cite{CHEN2019171}, the temperature dependence of the superfluid density of \LaCuNiAsO \ measured by TF-\msr \ is best described by the two-band model with the dominant $s$ wave superconducting energy gap larger for optimal doping, suggesting the coupling strength decreases with the increase of doping concentration. Both samples are classified between unusual superconductors and traditional BCS superconductors in the Uemura plot. More experimental evidence is required to investigate whether $d$-wave superconductivity exists in one of the bands. 
	
\begin{acknowledgments}
	
This work is based on experiments performed at the Swiss Muon Source S$\mu$S, Paul Scherrer Institute, Villigen, Switzerland, and TRIUMF, Vancouver, Canada. The research performed in this work was supported by the National Key Research and Development Program of China, No. 2022YFA1402203, the National Natural Science Foundations of China, No.~12174065 and No.51922105, and the Shanghai Municipal Science and Technology (Major Project Grant No.~2019SHZDZX01 and No.~20ZR1405300).
		
\end{acknowledgments}
	
%

\end{document}